\documentclass[prd,floatfix,twocolumn]{revtex4}

\usepackage{amsmath}  
\usepackage{amsfonts}  
\usepackage{graphicx}   

\newcommand{\be}{\begin{eqnarray}}
\newcommand{\ee}{\end{eqnarray}}

\newcommand{\vE}{{\bf E}}
\newcommand{\va}{{\bf a}}
\newcommand{\vd}{{\bf d}}
\newcommand{\vf}{{\bf f}}
\newcommand{\vp}{{\bf p}}
\newcommand{\vvr}{{\bf r}}
\newcommand{\vu}{{\bf u}}

\newcommand{\vx}{{\bf x}}

\newcommand{\ffF}{{\sf F}}
\newcommand{\ffU}{{\sf v}}

\newcommand{\ffp}{p}

\newcommand{\dtau}{\rm d\tau}
\newcommand{\dt}{{\rm d}t}

\newcommand{\dby}[2]{ \frac{{\rm d} #1}{{\rm d} #2}}
\newcommand{\pd}[2]{ \frac{\partial #1}{\partial #2}}

\newcommand{\Grad}{\mbox{\boldmath $\nabla$}}

\newcommand{\myfig}[2]
{\centerline{\resizebox{!}{#1\textwidth}{\includegraphics{#2}}}}

\begin{document}

\title{The non-existence of the self-accelerating dipole, and related questions}

\author{Andrew M. Steane}
\affiliation{Department of Atomic and Laser Physics, Clarendon Laboratory, Parks Road, Oxford OX1 3PU, England.}

\date{\today}

\begin{abstract}
We calculate the self-force of a constantly accelerating electric dipole, showing, in particular,
that classical electromagnetism does not predict that an electric dipole could
self-accelerate, nor could it levitate in a gravitational field. We also resolve
a paradox concerning the inertial mass of a longitudinally accelerating dipole, showing
that the combined system of dipole plus field can be assigned a well-defined energy-momentum
four-vector, so that the Principle of Relativity is satisfied.
We then present some general features of electromagnetic phenomena in
a reference frame described by the Rindler metric, showing in particular that
an observer fixed in a gravitational field described everywhere by the Rindler metric will
find any charged object supported in the gravitational field to possess
an electromagnetic self-force equal to that observed
by an inertial observer relative to which
the body undergoes rigid hyperbolic motion. It follows that the Principle of Equivalence
is satisfied by these systems.
\end{abstract}


\pacs{03.30.+p, 03.50.De,  04.20.-q, 04.40.Nr}



\maketitle

\section{Introduction}

In 1984 F. H. J. Cornish proposed that according to classical electromagnetism,
a sufficiently small or highly charged electric dipole (to be precise, a rigid body consisting of two 
point charges separated by a short rod) could undergo self-accelerated motion \cite{86Cornish}. That is, after being
placed in the right initial conditions, it would experience a self-force in the direction of its acceleration
that was sufficient to maintain the acceleration, without the need for any applied external force.
It follows that such a dipole could also self-levitate in a gravitational field.

This claim was accepted uncritically at the time \cite{86Griffiths}, and the argument continues to be
repeated \cite{99Petkov}.
We will show that the claim is wrong---but for interesting reasons. It turns out to be
an example of a more general phenomenon that has long been misunderstood, 
and continues to be widely misunderstood, namely the correct
treatment of equations of motion when self-force is non-negligible.

It has been known for over a century that classical electromagnetism has difficulties in treating point-like
charges \cite{90Rohrlich,98Jackson,71Landau,97Rohrlich,04Spohn}.
If a point-like particle with a finite charge could exist, then it would produce around itself an electromagnetic field
whose strength diverges near the particle and whose total energy is infinite. One might `live with' this
problem by adopting the concept of `renormalization', arguing that only energy differences are physically
relevant and, by use of a suitable procedure to regularize the divergent integrals, sensible predictions
could be obtained. However it turns out that this is not sufficient on its own, because it leads
to equations of motion that have pathological runaway solutions. 
Cornish was well aware of this background and merely drew attention to a previously unnoticed
but especially simple case. 

In the case of the rigid spherical shell, 
the pathological cases can be ruled out by insisting
that an entity of given charge and observed mass cannot have a radius below a certain
minimum \cite{83Schwinger,97Rohrlich,06Medina,10Griffiths}. This is connected to the fact
that no physical entity can have a negative mass---a simple
enough fact, but one which can be hidden when electromagnetic energy and momentum has to
be taken into account. We will show that the resolution in the case of the dipole is similar. 
The dipole case remains interesting, however, because the case for self-acceleration seems to be 
straightforward at first sight.

We also consider the fact that the electromagnetic self-force of a dipole depends on its
orientation with respect to its acceleration. This appears to imply the inertial mass of the dipole
depends on the orientation, which would present a difficulty with the 
Principle of Relativity and the Principle of Equivalence, 
because the field energy does not have such a dependence. Therefore it was
considered paradoxical \cite{83Griffiths,06Pinto,03Ori,04Ori}. We resolve this paradox
by appealing to the inertia of pressure. 

An alternative resolution was offered by Ori and Rosenthal \cite{03Ori,04Ori}, based on
a different, but well-motivated, definition of self-force also described by Pearle \cite{82Pearle}.
We reconcile the two approaches.

We also consider the case of a charged body at rest in an accelerating reference frame
in flat spacetime. This is the frame described by the Rindler metric; it describes the
simplest possible gravitational field (one which causes acceleration but not tidal effects). 
We present a general calculation of electromagnetic self-force this case. Our approach
to calculating the electromagnetic field agrees with several earlier 
treatments \cite{62Bradbury,63Rohrlich,04Eriksen}, but not, at first appearance,
with a recent calculation by Pinto \cite{06Pinto}. The difference is resolved by
considering what is meant by observation in or relative to an accelerating frame;
this influences the way forces acting at different positions should be summed or
compared. 

The paper is laid out as follows. Section \ref{s.dipole} treats the self-force of
an accelerating dipole. We first show that self-acceleration does not occur when the
properties of the dipole are restricted to physically possible values, and then we address the mass paradox
associated with the dependence of self-force on orientation. The analysis is tractable
when we model the dipole as two small spheres whose separation is large compared
to their radius. In order to address the complete problem it is necessary also to consider
the case where the spheres are close together; this is addressed by numerical calculations
in section \ref{s.numerical}. Section \ref{s.ori} presents the fact that there is more than
one way to consider what is the rate of change of momentum of an extended object,
owing to the relativity of simultaneity. Section \ref{s.grav} presents
the problem of electromagnetic self-force in the presence of gravity, in the simplest case. An exact
(General Relativistic) treatment turns out to be quite simple in the case of the Rindler
metric. Section \ref{s.conc} summarizes the conclusions.

\section{The accelerating dipole}  \label{s.dipole}

The `dipole' under consideration consists of a pair of charges $\pm q$ connected by
a short rod of proper length $d$ and undergoing rigid motion. 
By `rigid motion' we mean motion such that at each moment there is an inertial frame in 
which both charges are at rest, and their proper separation is constant,
i.e. the separation is the same in all successive instantaneous rest frames.
We need not assume that the rod is made of rigid material
(which would be impossible)---instead we assume that $d$ is the length
it adopts, in equilibrium, under the influence of the compressive forces from the two charges
as they attract one another, and any other external forces to which it may be subject, and we consider
a case where the external force is such that $d$ is constant.

In particular, we consider such a dipole undergoing motion at constant proper 
acceleration $a_0$ (``hyperbolic motion''), and in the first instance we treat
the case where the rod is aligned perpendicular to the acceleration.

\begin{figure}
\myfig{0.35}{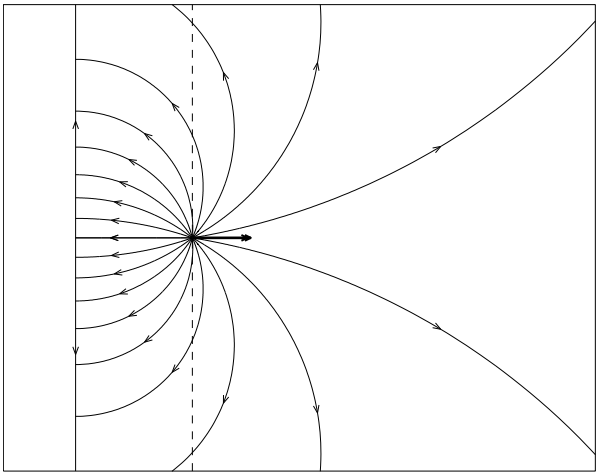}
\caption{Field-lines of the electric field in the instantaneous rest frame of a positively charged
small object undergoing constant proper acceleration in the $+x$ direction. A negative charge situated
anywhere on the dashed line
will experience a force with a component in the $+x$ direction, and will itself produce
an electric field similarly tending to accelerate the first charge.}
\label{f.fieldline}
\end{figure}

Figure \ref{f.fieldline} shows the lines of electric field from one of the charges, in the instantaneous
rest frame. For illustration, the field ${\bf E}_+$ of the positive charge is shown. 
Note that, at the location of the other (negative) charge, ${\bf E}_+$ is directed outwards and
somewhat in the direction opposed to the acceleration. Therefore, the force on the negative
charge, owing to the field of the positive charge, is inward and somewhat in the direction
along the acceleration. Similarly, the force on the positive charge, owing to the field of the
negative one, is also inward and somewhat in the direction along the acceleration. By forming
the sum of these two forces, one concludes that there is a net electromagnetic self-force
along the direction of the acceleration. This seems to suggest that this self-force could
provide the force required to make the dipole accelerate, and hence one would have a
self-accelerating dipole. 

We have presented this qualitative argument first, in order to show how natural the
suggestion of self-acceleration is in this case. Next we back it up with some quantitative
statements.

Let the motion be along $x$ and let the charges by separated by
a rod of fixed length $d$ aligned along $y$. The field due to each charge
at the position of the other has been obtained by Fulton and
Rohrlich\cite{60Fulton} \footnote{See \cite{12Steane}, section 8.2.5}. In the instantaneous rest frame it is given by
\be
|E_x| = \frac{4 q L^2}{4 \pi \epsilon_0 d (4 L^2 + d^2)^{3/2}}, \;\;\;\;
|E_y| = 2 |E_x| L / d                 \label{Exy}
\ee
where $L \equiv c^2/a_0$ and
the signs are such that the charges attract in the $y$ direction and each accelerates the
other in the $x$ direction. 

Each charge also experiences a self-force which can be treated
by using the Abraham-Lorentz-Dirac (ALD) equation. In order to do this, we first model each charge
as a small spherical shell of radius $R$, and then take the limit $R \ll d$. At small but finite $R$,
the electromagnetic self-four-force of such a shell is given by the ALD equation:
\be
\ffF_{\rm shell}  = \frac{2}{3}e^2\left( -\frac{\dot{\ffU}}{R} + \ddot{\ffU} 
- \dot{\ffU}^2 \ffU  + O(R)   \right) 
\label{ALD}
\ee
where we introduced $e^2 \equiv q^2/4\pi\epsilon_0$ to reduce clutter, the dot signifies
the derivative with respect to proper time, the four-vectors are displayed in index-free notation,
and we took $c=1$. 
If the shell is not exploding under the influence of its own electromagnetic
forces, then the material constituting it must be in tension. These internal stresses (Poincar\'e stresses)
also give rise to a self-four-force, discussed in appendix \ref{append1}, given by
\be
\ffF_{\rm P}  = \frac{1}{6}e^2 \frac{\dot{\ffU}}{R}   + O(R) .     \label{poincare}
\ee

In the case of
hyperbolic motion the second and third terms in Eq. (\ref{ALD}) are equal and opposite. Hence one finds that,
when $R \ll d$, the equation of motion for either charge of the dipole, when written 
in the instantaneous rest frame (and after reinstating $c$), is
\be
f_{\rm ext} + 
 \frac{4 e^2 L^2}{ d (4 L^2 + d^2)^{3/2}} 
-\frac{2 e^2}{3R c^2} a_0  + \frac{e^2}{6R c^2} a_0   = m_{00} a_0,   
\ee
where $m_{00}$ is the bare rest mass of the spherical shell in the absence of internal stress.

Let us introduce
\be
m_{\rm es} \equiv \frac{e^2}{2R c^2}
\ee
which is the total field energy of a spherical shell of charge that is permanently at rest (evaluated in
the rest frame). Then we have
\[
f_{\rm ext} + 
 \frac{4 e^2 L^2}{ d (4 L^2 + d^2)^{3/2}} 
-\frac{4}{3} m_{\rm es} a_0  + \frac{1}{3} m_{\rm es} a_0   = m_{00} a_0.
\]
It makes good physical sense to move the Poincar\'e stress term to the right hand side of the equation,
writing
\be
f_{\rm ext} + 
 \frac{4 e^2 L^2}{ d (4 L^2 + d^2)^{3/2}} 
-\frac{4}{3} m_{\rm es} a_0     = m_0 a_0,       \label{eqmotion}
\ee
where $m_0 \equiv m_{00} - m_{\rm es}/3$. This is good practice because $m_0$ is the inertial
mass of a well-defined physical
entity (the material of the shell, including its internal stresses) that is being acted upon by forces
external to it. This $m_0$ is often called the `bare mass'. It is
also customary to gather all the terms that are proportional to $a_0$, and
introduce $m \equiv m_0 + (4/3) m_{\rm es} = m_{00} + m_{\rm es}$.
$m$ is the mass that will be `observed', i.e. deduced from measurements of 
the acceleration of a single shell under given applied forces, if one chooses to
move all the inertial terms in the self-force to the right hand side of the equation of motion.

We want to know whether
(\ref{eqmotion}) has interesting solutions when $f_{\rm ext} = 0$.
After substituting $L = c^2/a_0$ and $f_{\rm ext}=0$, Eq. (\ref{eqmotion}) gives
a cubic equation for $a_0^2$. It has a single real solution for $a_0^2$, given by
\be
a_0^2 = \left(\frac{2c^2}{d}\right)^2 \left[ \left( \frac{e^2}{2 m c^2 d}\right)^{2/3} - 1 \right].
\label{a0Cornish}
\ee
This is the main result obtained by Cornish. One observes that for 
\be
d < \frac{e^2}{2 m c^2}          \label{dlimit}
\ee
one can have a solution of the equation of motion in which there is a constant acceleration
with no applied force. This is the surprising result whose validity we will question.

It is instructive to consider the force exerted by each charge on the other, i.e. the term involving $d$ in
Eq. (\ref{eqmotion}), also in terms of inertia. When $d \ll L$ this term is approximately
$e^2 /2Ld = \Delta m a_0$ where $\Delta m = e^2/2 d c^2$. Therefore the pair of spheres has
its total inertial mass reduced by 
\be
2 \Delta m = e^2/d c^2,             \label{massred}
\ee
which is precisely the `potential
energy' of a pair of point charges at rest, separated by $d$. Of course this `potential energy' is
really field energy: it is the amount by which the field energy is smaller, when the charges are
brought to separation $d$, compared to when they are far apart, in the case $R \ll d$. 
One may then observe that if $\Delta m > m$ then one would have a negative total effective mass,
and therefore self-acceleration. Thus one deduces the condition (\ref{dlimit}) again.

Since energy and momentum are exactly conserved in the interaction between
particles and fields in classical electromagnetism, the existence of self-accelerated solutions has sometimes
been interpreted, somewhat vaguely, as a way of drawing on the infinite reserves of energy to be
found in the electromagnetic field near a point-like particle.  However, such an
argument will not work, because we don't need $R$ to be zero, only small, so the field
energy is finite. It begins to look as if energy-momentum conservation is breaking down.

In fact there is no such conclusion.
The problem with the result is that the condition (\ref{dlimit}) cannot be satisfied
if $m_0 \ge 0$. For then one has $m \ge (4/3) m_{\rm es}$ so
\be
\frac{e^2}{2 m c^2} \le \frac{3}{4} R.
\ee
Hence if we model the dipole as two small spherical shells, as above, then the 
condition for the self-accelerated
solution is that the centres of the shells are separated by substantially less than twice
their radius. But this implies that they overlap and therefore the calculation is invalid.

If one admits $m_0 \le 0$ then it should not surprise us that self-acceleration could
occur. As the matter with $m_0 < 0$ accelerates, its 
kinetic energy gets more and more negative, and a corresponding
energy goes into the fields, since energy is conserved overall. But the whole situation
remains unphysical.

\subsection{Longitudinal dipole: resolution of mass paradox}

The self-force for the case of a longitudinally-accelerating dipole is equally easy to extract using the
equations for the field of an accelerating point charge.
We consider a pair of point charges undergoing constant proper acceleration along the
$x$-axis. The condition for rigid motion (i.e. constant proper separation) is
that the charges have proper accelerations given
by\cite{12Steane,13Steane} $a_i = c^2 / x_i$ where $x_i$ is the location of the $i$'th charge in the
instantaneous rest frame. It follows that if the particles are separated by a rod of proper length
$d$, and the centre of the rod has proper acceleration $a_0$, then the proper
accelerations of the two particles are given by
\[
\frac{a_0}{1 \pm a_0 d/2 c^2},
\]
the trailing particle having the higher acceleration. 

The electric field at $x_1$ on the $x$-axis due to an accelerating point charge $q_2$ located
at $x_2$, in the instantaneous rest frame, is\footnote{See \cite{60Fulton} or Eq. (8.74) of \cite{12Steane}}
\be
E^{(1,2)} = s \frac{ 4 q_2 x_2^2}{4 \pi \epsilon_0 (x_2^2 - x_1^2)^2},         \label{E12}
\ee
where $s$ is the sign of $(x_1-x_2)$
and the origin has been located such that $x_2 = c^2/a_2$. By interchanging the labels
one finds that the total self-force (ignoring the self-force of
each charge on itself) is
\be
q_1 E^{(1,2)} + q_2 E^{(2,1)} = \frac{4 q_1 q_2}{4 \pi \epsilon_0 (x_1^2 - x_2^2)}
= \frac{2 e^2}{d c^2} a_0.         \label{flong}
\ee
The calculation is exact in the limit $d \gg R$; it gives the self-force in the instantaneous rest frame.
The final form on the right hand side of Eq. (\ref{flong})
suggests an interpretation in terms of mass, and it shows that in this case
the inertial mass reduction is by twice what one might expect
(for example it is twice that observed in
the transverse case, Eq. (\ref{massred})).
This is the paradox noted by Giffiths and Owen\cite{83Griffiths}
and taken up by Pinto\cite{06Pinto} (see also \cite{94RoaNeri}).

The paradox is not that the self-force depends on orientation, but with reconciling this fact with
energy considerations. To lowest order in $a_0$ the field energy does not depend on orientation: it
is given by $-e^2/d$ plus a contribution that is independent of the positions of the charges. Therefore
it appears as if the energy and momentum of the complete system (dipole plus field) will not be able
to form a four-vector at all orientations. This would violate basic principles of Special
Relativity. It would also violate the Principle of Equivalence, since the
passive gravitational mass of the complete system (matter plus field)
is determined by the energy (divided by $\gamma c^2$) whereas
inertial mass is determined by the momentum (divided by $\gamma v$).

Since energy-momentum is exactly conserved in classical electromagnetism, we can be sure that Eq. (\ref{flong})
gives the (negative of the) rate of change of the field momentum. To be precise, it matches the part of
the field momentum associated with cross terms. This was checked to first order approximation 
by Giffiths and Owen, and we can rely on the consistency of the theory to be assured that it will be true
exactly. The only mystery is that this momentum is not matching up with the field energy in
the appropriate way: we have a `mysterious' factor 2.

The resolution is as follows.

This `2 problem' is just like the famous `4/3 problem' in the treatment of a charged sphere, 
and it can be understood in the same way. We have
to take into account the pressure in the rod \footnote{Griffiths and Szeto \cite{78Griffiths} make
this same observation without providing details; however in \cite{83Griffiths} it was overlooked.}.
There is no choice about this: the physical system
could be realized by placing two real, physical charged spheres at the end of a literal rod, and such a
rod will certainly thus be placed in compression. The issue does not arise in the transverse case because
in that case the pressure forces in the rod are transverse to the motion. In general, however,
the pressure does influence the dynamics. 
One can think of this either in terms of `hidden momentum'
\cite{12Steane} or in terms of the contribution of pressure to inertia (c.f. appendix \ref{append1}).
When one calculates the contribution of the pressure to the inertia exhibited by the rod, one finds its
inertia tensor is not isotropic.

Consider the two charged spheres separated by a rod lying along the direction of acceleration,
which we continue to take as the $x$-direction. Let us treat the material of the rod as an ideal
fluid (imagine a fluid-filled tube with the charged spheres attached to pistons at each end). In the
instantaneous rest frame, the pressure $p$ in the fluid will obey the relativistic Navier-Stokes
equation, which in the instantaneous rest frame takes the form
\be
\left(\rho_0 + \frac{p}{c^2} \right) \frac{{\rm D}\vu}{{\rm D}t} = - \Grad p.  \label{NavStok}
\ee
For positive pressure we can take the mass density of the fluid ($\rho_0$) to be negligible, and
for the rigid hyperbolic motion under consideration, each part of the fluid has a proper
acceleration given by ${\rm D}u / {\rm D}t = c^2/x$. Hence the solution of the Navier-Stokes
equation is 
\be
p \propto \frac{1}{x}    .      \label{pfromx}
\ee
If the cross-section of each piston is A then the equations of motion of the two charged spheres are
\be
f_{\rm ext}^{(1)} + f_1 - p_1 A &=& m a_1   \nonumber  \\
f_{\rm ext}^{(2)} -  f_2 + p_2 A &=& m a_2              \label{eqmotions}
\ee
where $m$ is the observed mass of each sphere,
$f_{\rm ext}^{(i)}$ is
the external force on sphere $i$, $p_i$ are the pressures at the two ends of the tube, $a_i = c^2/x_i$
are the accelerations of the spheres, and $f_ i$ is the magnitude of the force on sphere $i$ owing to the field of the
other sphere, given by eqn (\ref{E12}):
\be
f_1 = \frac{e^2}{d^2 L^2} x_2^2, \;\;\;\;
f_2 = \frac{e^2}{d^2 L^2} x_1^2.
\ee
where the two spheres are centred at $x_1 = L - d/2$, $x_2 = L + d/2$.
By using (\ref{pfromx}) in (\ref{eqmotions}) we find
\be
\frac{f_1 + f_{\rm ext}^{(1)} - m a_1}{f_2 - f_{\rm ext}^{(2)} + m a_2} = \frac{x_2}{x_1}
\ee
Hence
\be
f_{\rm ext}^{(1)} x_1 + f_{\rm ext}^{(2)} x_2 = 2 m c^2 - ( f_1 x_1 -  f_2 x_2).  \label{fxfx}
\ee
In general there is no compelling reason why the external forces on the two spheres need be equal.
However, if we suppose that they are then we find that the total external force is related to the acceleration
of the centre of the rod, $a_0 = c^2/L$, by
\be
2 f_{\rm ext} = 2 m a_0 - \frac{e^2}{d c^2} \left(1 - \frac{a_0^2 d^2 }{4 c^4} \right) a_0  \label{selflong}
\ee
Hence in the limit $a_0 d \ll c^2$ we find that the self force, after taking internal pressure into
account, matches the result for transverse orientation of the dipole, eqn (\ref{massred}). This
confirms that the behaviour of the momentum is consistent with the behaviour of the energy,
for small dipoles or small accelerations, for these two orientations of the dipole, and we will
show it for all orientations in the next section.

For the case where the external force varies as $f_{\rm ext}^{(i)} \propto 1/x_i$, the left hand side
of (\ref{fxfx}) evaluates to $2 f_{\rm ext} L$, where $f_{\rm ext}$ is now the value
of the force for $x=L$, and one obtains (\ref{selflong}) again. In this case the total external
force is not $2 f_{\rm ext}$  but $2 f_{\rm ext} (1-d^2/4L^2)^{-1}$. 

One may also interpret the physical picture in terms of `hidden momentum', as follows.
`Hidden momentum' is momentum associated with energy transport through
the body \cite{12Steane}. As the rod accelerates, the hidden momentum continually
increases. This increases the inertia of the rod by
the integral of the force along the length of the rod\footnote{\ldots for uniform acceleration. More
generally, one must integrate eqn (\ref{NavStok}).}, which is $e^2/d$. This happens to
be equal to the electrostatic field energy, but it is located in a completely different physical system,
namely the material of the rod. Once this energy is added to the electrostatic field energy, we get
a complete system (charges plus rod plus surrounding field) which can be treated as isolated and assigned
a four-momentum. In particular, we find that the external force required to accelerate the dipole
(or to keep it at a fixed location in a gravitational field) does not depend on the orientation of
the dipole, for small $a_0$. In the longitudinal case, the dipole `pulls itself along' by its own electromagnetic
forces more than in the transverse case; however it has to do this in order to provide the
hidden momentum associated with its internal pressure forces as 
well as its ordinary momentum, 
with the net result that its overall tendency to resist acceleration by outside forces is the
same in the longitudinal as in the transverse case. 

The condition for self-acceleration in the longitudinal case, after taking hidden momentum into
account, is the same as for the transverse case, namely condition (\ref{dlimit}), but as before
this is outside the range of validity of the calculation if we insist that bare mass is non-negative.
In either case, longitudinal or transverse, although we do not expect self-acceleration, we do expect
that a dipole will be observably lighter than an object otherwise similar but with two
charges of the same sign. The expected difference in observed mass between the dipole and the dumbbell
is twice Eq. (\ref{massred}), i.e. $2 e^2/d c^2$.

All the above is valid for $a_0 d \ll c^2$. More generally, the self force given by eqn (\ref{selflong})
does not exactly match that given by (\ref{eqmotion}). Both of these equations have been
derived without restriction on the value of $a_0$, except for the restriction imposed by
the horizon at $x=0$, namely $d < 2L$ so $a_0 d < 2 c^2$, and we are still assuming
$R \ll d$. However, comparing the dipole at one orientation with the dipole at another is non-trivial
once the acceleration is substantial, because it is no longer clear what value should be considered
`the acceleration of the dipole' when different parts have different accelerations (the longitudinal
case), nor is it easy to locate the centroid of the field energy distribution.
The following argument shows that a difference in self-force between the transverse and
longitudinal case is expected at $O(a_0^3)$. The situation is comparable to the case of an
object fixed in a gravitational field whose strength varies as $g \propto 1 / x$. Then for
a prolate object of length $d$ at height $L$, the total 
gravitational force when it is oriented vertically exceeds that when it
is oriented horizontally by an amount of order $(d/L)^2 f$, where $f$ is the gravitational force
in the horizontal case. We can apply this fact to the mass distribution associated
with the field energy. The electromagnetic contribution to the mass is of order $e^2/d c^2$,
and this mass is mostly concentrated in a prolate region of size approximately $d$.
Therefore we expect an orientation-dependent contribution to the 
electromagnetic self-force of order
\be
\left( \frac{d}{L} \right)^2 \frac{e^2 a_0}{d c^2} = \frac{e^2 d a_0^3}{c^6}.  \label{acubed}
\ee

\subsection{Dipole at arbitrary orientation}

We presented the transverse and longitudinal cases in detail in order to get clarity about
the underlying physical mechanisms, and because it permits some simple exact results (in the limit
$d \gg R$) to be exhibited, such as eqns (\ref{eqmotion}), (\ref{a0Cornish}), (\ref{flong})
and (\ref{selflong}). For a dipole
at arbitrary orientation to its acceleration, we shall treat the problem to first order in
the proper acceleration. The electric field produced by the first charge
at the second is given by the standard expressions for the electric field of a charge
in hyperbolic motion, and by expanding to first order in $a_0$ one finds
\be
\vE^{(2,1)} = \frac{q}{4\pi\epsilon_0}\left[ \frac{\hat{\vd}}{d^2}
-  \frac{\va_0 + (\va_0 \cdot \hat{\vd}) \hat{\vd}}{2c^2 d} \right]
+ O(a_0^2)    \label{Egenangle}
\ee
where $\hat{\vd}$ is a unit vector in the direction from the first charge to the
second. The total electromagnetic self-force of the dipole is therefore, to $O(a_0)$,
\be
\vf_{\rm self}^{\rm (e.m.)} \simeq \frac{e^2}{c^2 d}\left(\va_0 + (\va_0 \cdot \hat{\vd}) \hat{\vd} \right)
+ 2\vf_{\rm sphere}^{\rm (e.m.)}           \label{fselfemangle}
\ee
where $\vf_{\rm sphere}$ is the force of each charged sphere on itself (this is in the direction
opposite to $\va_0$). The pressure force
in the rod varies monotonically from one end to the other, but to $O(a_0)$ it
is sufficient to use the average along the rod, which (again to $O(a_0)$)
is equal to the average of the magnitudes of the two forces on the ends, i.e. 
$(e^2/d^3) \vd$. Hence the rate of change of hidden momentum is
\be
&& \frac{e^2 }{c^2 d^3}  \int_{-d/2}^{d/2} (\vd \cdot \va_0)\hat{\vd}  \, {\rm d} s  \nonumber \\
&=& \frac{e^2}{c^2 d^3} (\vd \cdot \va_0) \vd  ,             \label{rodforce}
\ee
where, in the integral, $s$ is distance along the rod.
The hidden momentum may be accounted for by bringing it to the other side of the equation
of motion and regarding it as a contributer to the self force. Hence, by subtracting (\ref{rodforce})
from (\ref{fselfemangle}), and also including the effect of internal stresses in the spheres,
we find the total self-force of the system is
\be
\vf_{\rm self} \simeq \frac{e^2}{c^2 d} \va_0 + 2\vf_{\rm sphere} .          \label{fselftotal}
\ee
Hence the resistance to acceleration by external forces is independent of the orientation of the rod,
for all angles, to first order in $a_0$, when $d \gg R$. The whole situation is closely
related to the Trouton-Noble experiment \cite{45Page}.

Historically the electromagnetic contribution to the mass of extended entities such as
atoms and molecules has been considered to be of purely theoretical interest, being
too small (of order $10^{-10}$ of the rest mass) to be observed experimentally. However,
modern mass comparison techniques using  ions trapped in Penning traps can achieve
the required sensitivity \cite{04Rainville}. It would be interesting, for example, to confirm that
the inertial mass of a polar molecule such as lithium hydride is independent of its orientation. This
would show that the quantum mechanical source of the internal pressure in the molecule, namely
zero point energy when an electron is confined to a small region, 
gives rise to the requisite hidden momentum as Special Relativity says it must.

\subsection{Dipole with large spheres}  \label{s.numerical}

\begin{figure}
\myfig{0.15}{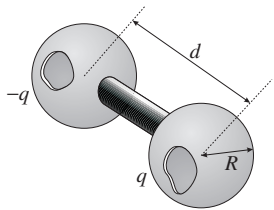}
\caption{A pair of charged spherical shells separated by a rigid rod.}
\label{f.dipole}
\end{figure}

If we avoid the assumption of negative bare mass, then the 
remaining possibility, if we are searching for self-accelerating solutions,
is to suggest that there might exist
some charge distribution which gives a net electromagnetic self-force in the direction
of the acceleration, even when the inertial terms are included.
Given that the fields
around an accelerating charge tend to retard any like charge moving alongside
the first one, the most promising distribution would appear to be a dipole-like
form, but made of a pair of larger spheres, so as to reduce $m_{\rm es}$
as much as possible without greatly changing the force exerted by each sphere
on the other. Therefore let us consider two oppositely charged spherical
shells, having total charges $\pm q$, radius $R$, with their centres separated by
$d$ (figure \ref{f.dipole}). The calculation in the previous section already applies to this
`dipole' when $d \gg R$, but we would like to find out whether the 
case $d \simeq 2R$ (or indeed $d < 2R$, i.e. intersecting spheres)
can yield self-acceleration.

The electromagnetic self-force of a single spherical shell of charge undergoing hyperbolic motion
has been calculated exactly \cite{13Steane}. In the instantaneous rest frame, it is
\be
f_{\rm shell} &=& \frac{2 e^2}{R c^2} a_0 \sum_{n=0}^\infty \frac{ (R a_0 / c^2)^{2n} }
{ (2n-1)(2n+1)^2 (2n+3) }          \label{fselfexact}     \\
& \simeq & \frac{e^2}{L R} \left[ -\frac{2}{3} + \frac{2}{45} \left( \frac{R}{L} \right)^2 
+ \frac{2}{525} \left( \frac{R}{L} \right)^4 + \ldots \right]  \nonumber
\ee
and the condition that the sphere can maintain its proper size and shape is $R < L$.
This shows that the further terms in the power series expansion do lower the absolute magnitude of the
self-force of the shell, but one finds this reduction is not by enough to allow a self-accelerating dipole,
as we now show.
The force $f_{\rm dip}$ of each shell on the other is in the forward direction. It can be estimated for
$d \gg R$ by using Eq. (\ref{Exy}), which gives
\be
f_{\rm dip} \simeq  \frac{e^2}{L d} \left[ \frac{1}{2} - \frac{3}{16} \left( \frac{d}{L} \right)^2 
+ \frac{15}{256} \left( \frac{d}{L} \right)^4 + \ldots \right].
\ee
For example, at $R=L,\;d=2R$ (i.e. large, touching spheres)
one finds $f_{\rm shell} = -(\pi^2/16) e^2/L^2$ and $f_{\rm dip} \simeq (8 \sqrt{2})^{-1} e^2/L^2
\simeq 0.14 |f_{\rm shell}|$.

In order to confirm this conclusion we need to replace the rough estimate for $f_{\rm dip}$ by
a more accurate value. We did this by numerical integration, as described in appendix \ref{append2}.
Figure \ref{f.ftot} shows the results.  For all values of $d/R$ and $R/L$ we find that
the total self-force opposes the acceleration.

\begin{figure}
\myfig{0.33}{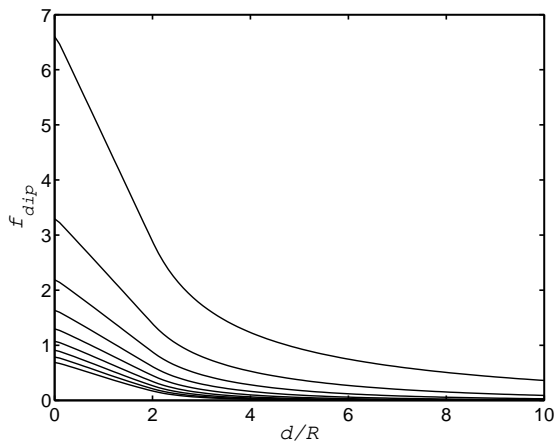}
\caption{The contribution $f_{\rm dip}$ to the self-force of a pair of oppositely-charged spherical shells
of radius $R$ with centres separated by $d$ and undergoing rigid hyperbolic motion in the transverse
direction.The force is shown in 
units of $e^2/L^2$, for nine equispaced values of $R$ between $0.1\,L$ and $0.9\,L$.
The total self-force of the pair of spheres is $f = 2(f_{\rm dip} + f_{\rm shell})$ where $f_{\rm shell}$
is given by Eq. \protect\ref{fselfexact}. As $d \rightarrow 0$ one finds that $f_{\rm dip} \rightarrow -f_{\rm shell}$
(see text) so $f \rightarrow 0$. The spheres are just touching when $d/R=2$. At $d \gg R$, $f_{\rm dip}$ is
independent of $R$ and is given by the $d$-dependent term in Eq. \protect\ref{eqmotion}.}
\label{f.ftot}
\end{figure}

\begin{figure}
\myfig{0.33}{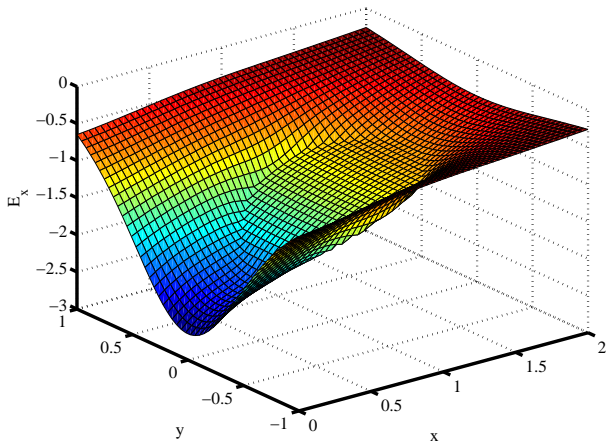}
\caption{The $x$-component of the field $\bar{\vE}_{\rm shell}$, which is the
field of an accelerating sphere with a radially symmetric contribution removed, so
as to leave a continuous function whose integral can be used to calculate
the self-force for the case of a transversely oriented dipole. The figure 
shows the case $L=1$, $R=1/2$ for illustration.
Note that $\left|\bar{\vE}_{{\rm shell},x} \right|$ falls monotonically with $y$
for points outside the shell.}
\label{f.Ex}
\end{figure}

One can prove that the self-force vanishes in the limit $d \rightarrow 0$ as follows.
Consider the field
$\vE_{\rm shell}$ due to a single charged shell. It is
discontinuous at the edge of the shell by $(\sigma / \epsilon_0) \hat{\vvr}$ 
where $\sigma=q/4\pi R^2$ is the surface charge density and $\hat{\vvr}$ is a unit vector
in the direction radially outwards from the centre of the shell.
Therefore the field
$\bar{\vE}_{\rm shell} \equiv \vE_{\rm shell} - \sigma H(r/R)  \hat{\vvr} / \epsilon_0 r^2$
is continuous, where $H(x)$ is the Heaviside step function.  The fields
$\bar{\vE}_{\rm shell}$ and $\vE_{\rm shell}$ differ by a field that exerts no
net force in the $x$-direction on any charge distribution that is symmetric about $x=L$. 
Therefore we can use
either of them for the purpose of calculating the self-force of the transversely-oriented
dipole. By using $\bar{\vE}_{\rm shell}$ we eliminate the discontinuity; this allows
the rest of the argument to proceed.
Now consider the two contributions $f_{\rm shell}$ and $f_{\rm dip}$. Both
may be calculated by integrating $\bar{\vE}_{\rm shell}$ over a spherical
charge distribution. The two charge distributions in question have opposite sign
and infinitesimally different locations in the limit $d \rightarrow 0$. Therefore in that
limit one must find $f_{\rm shell} = -f_{\rm dip}$.
This is expected since in this limit the fields produced
by the two shells cancel. For $d > 0$ one can see from the overall form
of the integrand (figure \ref{f.Ex}) that $f_{\rm dip}$ must fall monotonically
as $d$ increases. Therefore in this physical system
the forward force arising from the presence of opposite charges
can just approach the backward force arising from the presence of like charges, but cannot
exceed it and thus produce self-acceleration.

The above considerations for $d \rightarrow 0$ will also apply if we model the
pair of charged objects using some other shape or distribution of charge. This suggests
that the overall conclusion, that the total self-force never points in the direction
of acceleration, will hold true more generally. Further calculations would be needed
to confirm this.

\section{Alternative definitions of self-force}  \label{s.ori}


So far we have presented the self-force by adopting the policy of selecting a reference
frame at the outset (the instantaneous rest frame) and summing the 3-forces acting
simultaneously in this frame. This is a valid method. However, owing to the relativity of simultaneity, it
is not the only one that may be regarded as legitimate and useful. 

Consider a composite object that can be decomposed into a set of discrete entities $i$.
The total 4-momentum of the composite object is
\be
\ffp^\mu_{\rm tot}(\tau_c,\chi) = \sum_{i}  \ffp_i^\mu \left(\tau_{i,\chi}\right)
\ee
where $\chi$ denotes a spacelike hypersurface,
$\tau_{i,\chi}$ is the proper time on the $i$'th worldline when that worldline intersects $\chi$,
and $\tau_c$ is the proper time on some reference worldline (e.g. the worldline of the centroid).
In other words, $\chi$ is the hypersurface on which the individual 4-momenta $\ffp_i^\mu$ are evaluated
in order to form the sum.
Typically, one picks a spacelike hyperplane (so that the events $\{i\}_\chi$ are
simultaneous in some frame). If the composite object is isolated then the result does not depend on
the choice of hyperplace \cite{12Steane}. If it is not isolated, which is the case for
any calculation of self-force (since then the object in question is being pushed or pulled by its own
electromagnetic field and by an external force), then $\ffp^\mu_{\rm tot}$ does depend on $\chi$.
For an object whose motion is rigid---that is, its motion is such that at any given event on
the world-tube there is a reference frame in which all parts of
the object are at rest, and at the same proper distances---a natural choice of~$\chi$ is
the hyperplace of simultaneity for the instantaneous rest frame at the given~$\tau_c$.

Suppose each discrete entity in the composite object
experiences a four-force $\ffF^\mu_i = {\rm d}\ffp_i^\mu / \dtau_i$.
Having established a definition of $\ffp^\mu_{\rm tot}$ at one instant, one may take an
interest in the rate of change of this quantity:
\be
\dby{\ffp^\mu_{\rm tot}}{\tau_c} = \lim_{\delta\tau_c \rightarrow 0}
\frac{\ffp^\mu_{\rm tot}(\tau_c + \delta\tau_c, \chi + \delta\chi) - \ffp^\mu_{\rm tot}(\tau_c,\chi)}{\delta\tau_c}
\label{defforce}
\ee
where we have assumed a one-to-one correspondence between $\chi$ and $\tau_c$, such that 
$\delta\chi \rightarrow 0$ as $\delta\tau_c \rightarrow 0$. 

The result (\ref{defforce}) depends on what choice is made for the hyperplane $\chi + \delta \chi$. 
So far in this paper we have adopted the instantaneous rest frame in order to pick $\chi$,
and the method of summing three-forces acting simultaneously in that frame amounts to choosing 
for $\chi + \delta \chi$ a hyperplane parallel to $\chi$ and separated from it
by a time $\delta t$ in the given frame. The result for the spatial part of 
${\rm d}\ffp^\mu_{\rm tot}/\dtau_c$ is
\be
\dby{\vp_{\rm tot}}{t} = \sum_i \dby{ \vp_i}{t}
\ee
where the quantities ${\rm d}\vp_i/{\rm d}t$ are evaluated on the hyperplane $\chi$.
For any given worldline we have $\dtau_i/\dt = 1$ in
the instantaneous rest frame, hence we may also write
\be
\dby{\ffp^\mu_{\rm tot}}{\tau_c} = \sum_i \dby{ \ffp^\mu_i}{\tau_i}.   \label{dptotA}
\ee
Another interesting choice for $\chi + \delta \chi$ is a hyperplane of simultaneity
for the new instantaneous rest frame at
$\tau + \delta \tau$. For an accelerating object this is not parallel to $\chi$, and one has
\be
\dby{\ffp^\mu_{\rm tot}}{\tau_c} &=& \lim_{\delta\tau_c \rightarrow 0}\sum_i \frac{\ffp^\mu_i(\tau_c + \delta\tau_i) -   \ffp^\mu_i(\tau_c)}{\delta\tau_c}    \label{sumchi}   \\
&=&  \sum_i \dby{\ffp^\mu_i}{\tau_i}  \dby{\tau_i}{\tau_c}   \label{newtot}
\ee
where in the sum in (\ref{sumchi}), each $\delta\tau_i$ is the proper time elapsed
on the $i$'th
wordline between the intersections of that worldine with $\chi$ and $\chi+\delta\chi$,
and in (\ref{newtot}) the quantities ${\rm d}\ffp^\mu_i/\dtau_c$
and $\dtau_i / \dtau_c$ are evaluated on the hyperplane $\chi$. We now have two
{\em different} definitions of the rate of change of the total momentum
for the composite body: eqn (\ref{dptotA}) is not the same as eqn (\ref{newtot}).
Hence the phrase `the self force' is ambiguous until one has specified which
definition is adopted.

Ori and Rosenthal \cite{03Ori,04Ori}, following Pearle \cite{82Pearle},
have described the approach using (\ref{newtot}).
This approach has the advantage that for rigid motion, the
internal forces cancel and the electromagnetic self force one obtains
is independent of the shape or orientation of the composite object. However
one should not ignore the internal forces altogether, and indeed in
the approach using (\ref{dptotA}) they play an important role, as we
have shown. (The statements in \cite{03Ori,04Ori} suggesting the inadmissability
of (\ref{dptotA}) are largely mistaken because they fail to take into account
the fact that the internal stress tensor need not be spherically symmetric).

\section{Self-force in a gravitational field} \label{s.grav}

The general problem of self-force in a gravitational field is rich and subtle, for recent reviews
see \cite{11Poisson,12Isoyama}. Here we consider only the case of a charged body
held fixed in a spacetime described
by the Rindler metric. This metric is appropriate to a uniformly accelerating reference
frame in flat spacetime. Obviously, this case does not show the
quintessential gravitational phenomena that are associated with curvature and tidal forces. 
However the uniformly accelerating reference frame is an important basic
case that can be used to explore phenomena that are associated purely with 
a spatial dependence of proper time, in the absence of spacetime
curvature. It is also very useful for gaining physical insight.

We shall be concerned with the purely electromagnetic force 
which includes a divergent part (in the limit of point-like objects) and a non-divergent part
commonly called radiation reaction.
The gravitationally-induced self-force $f_G$ discussed in \cite{79Vilenkin} vanishes
in flat spacetime and therefore we shall not be concerned with it (even though it may
dominate the radiation reaction in gravitational problems of
practical interest.)

Pinto\cite{06Pinto} has presented a calculation of the field of a point charge in a reference frame
described by the Rindler metric, by developing a formula for electric potential in
the Rindler frame and evaluating its gradient. He thus finds that the electromagnetic self-force 
for a dipole is independent of orientation, to lowest order in the acceleration.
Previously the electric field of a point charge in the constantly accelerating frame
was obtained by several workers \cite{62Bradbury,63Rohrlich,04Eriksen} using another
method, namely to start with the field tensor in Minkowski space and then transform it; see also
\cite{09Padmanabhan}. The field thus obtained differs from Pinto's, and gives a self-force for
a dipole that depends on orientation. We shall show that these differences arise from
the difference between definitions (\ref{dptotA}) and (\ref{newtot}). 

Before considering the point charge, we examine the electromagnetic field in the Rindler
frame in general. This will permit some observations more general than those
given by Bradbury or Rohrlich, and we will
bring out an interesting aspect not explicitly indicated by anyone. The derivation is quite simple.

We consider a region of flat spacetime. The region is mapped by a coordinate system ($T,X,y,z$) 
describing an inertial frame (one whose metric is Minkowskian), and also by another coordinate system
$(\theta,h,y,z)$ related to the first by
\be
\theta = \tanh^{-1}(T/X), \;\;\;\;\;\;\; h = \sqrt{X^2 - T^2}
\ee
in the region where $h$ is real and positive (we will not need to consider the rest of spacetime). 
The metric for this second system is the Rindler metric, $g_{a'b'} = {\rm diag}(-h^2,1,1,1)$.
Any point fixed in the second system is undergoing hyperbolic motion relative to the first, with
constant proper acceleration $1/h$ (we take $c=1$ throughout this section). One can see
immediately from the metric that the second system is static, i.e. the set of points at given $(h,y,z)$
form a rigid lattice with fixed proper distances between them: it is the `constantly accelerating
reference frame' in flat spacetime\cite{12Steane,06Rindler,04Eriksen}. 

The coordinate transformation matrix is
\be
\Lambda^{a}_{\;a'} \equiv \pd{x^{a}}{x^{a'}}  = \left( \begin{array}{cccc} 
\frac{1}{h}\cosh\theta & -\frac{1}{h}\sinh\theta & 0 & 0 \\ 
-\sinh\theta & \cosh\theta & 0 & 0 \\ 0 & 0 & 1 & 0 \\ 0 & 0 & & 1 \end{array} \right)
\ee
in which the primed (unprimed) indices correspond to the Minkowski (Rindler) coordinate system. Note the
similarity with the Lorentz transformation.

To calculate the electromagnetic effects we start in the first coordinate system and use
Maxwell's equations in flat spacetime to find the fields in the standard way. This means that
we neglect the effect of this electromagnetic field on the spacetime curvature; thus we neglect
some nonlinear effects which are negligible in the limit of weak fields. We thus find the
field tensor $F^{a'b'}$ in the Minkowski coordinate system. Since this transforms as an ordinary tensor,
we may immediately find its form in the Rindler system, given by $F^{ab} = \Lambda^{a}_{\; a'}
\Lambda^{b}_{\; b'} F^{a'b'}$. The result of this easy calculation is that the tensor transforms just
like it would under a Lorentz transformation, except that the first row and column (i.e. $F^{0b}$
and $F^{a0}$) pick up an additional factor $1/h$. Upon pre- and post-multiplying by $g_{ab}$,
which introduces a factor $h^2$, we find that the covariant 
form $F_{ab}$ has first row and column multiplied by $h$, compared to
a Lorentz-transformed version of $F_{a'b'}$.

To calculate the electromagnetic force on a charged particle in this field, we use
\be
\dby{p_{a}^{\rm (EM)}}{\tau} = q F_{a \lambda} \dby{x^{\lambda}}{\tau}  \label{dpEM}
\ee
where $p_{a}$ is 4-momentum and the superscript (EM) signifies that we are only
writing down the contribution from electromagnetic effects. (Note, however, that we shall introduce
another definition of the `electromagnetic force' shortly).
For example, consider a particle fixed at height $h$ in the Rindler frame. Its worldline in the Minkowski
frame is
$x^2 - t^2 = h^2$ and therefore its 4-velocity in the Minkowski frame is $u^{a'}=(\cosh \theta,\sinh\theta,0,0)$.
Upon transforming we find\footnote{This statement could
also be derived directly from the metric, using $u_{\lambda} u^{\lambda} = -1$.} 
$u^{a} = \Lambda^{a}_{\; a} u^a = (1/h,0,0,0)$. 
The factor $1/h$ in the 4-velocity exactly cancels the factor
$h$ in the first row of the field tensor, and we obtain
\be
\dby{p_{a}^{\rm (EM)}}{\tau} = \dby{p_{\bar{a}}^{\rm (EM)}}{\tau}.  \label{sameforce}
\ee
where the barred coordinates refer to the inertial frame obtained by Lorentz boost from
the original $(t,x,y,z)$ frame. Equation (\ref{sameforce}) asserts that
{\em for {\em any} given electromagnetic field in flat spacetime, the components of the
electromagnetic 4-force on a particle fixed in the Rindler frame, expressed in the coordinate system of that
frame, are the same as those of the electromagnetic 4-force on that same particle, expressed in the coordinate
system of a Minkowski frame relative to which the Rindler frame is momentarily at rest}.
Informally, one may say that, when calculating observations made by an observer
at rest in the constantly accelerating reference
frame, we don't need to worry about the general covariance of electromagnetism: just
Lorentz-boost to the instantaneous rest frame, and you will find the correct 4-force.

In the case of the self-force of any arbitrary charge distribution undergoing rigid acceleration,
the consequence of the above general statement is especially simple: the electromagnetic
self-force is the same in the Rindler frame as in the Minkowski instantaneous rest frame.
That is to say, the self-force as defined by (\ref{dptotA}) will agree in the two frames, and the
self-force defined by (\ref{newtot}) will also agree in the two frames. 

It follows that all our previous statements about electromagnetic
forces on an accelerating dipole also apply to electromagnetic forces on a dipole fixed in 
a gravitational field described by the Rindler metric. In particular, if we adopt the definition
(\ref{dptotA}) then the electromagnetic
self-force is larger when the dipole is oriented along the gravitational field
then when it is oriented transverse to the gravitational field. Indeed, in view of the pressure forces
in the rod, this must be the case if the Principle of Equivalence is to be satisfied. Having
taken the internal pressure into account, the net result is that the weight of a dipole is independent
of its orientation (assuming that it is in mechanical equilibrium). However, in the case
of the uniformly accelerating reference frame, (\ref{dptotA}) is not the most natural
definition of the total force on an extended object. Rather, (\ref{newtot}) is more natural.
If we adopt that definition then we find the electromagnetic self-force is independent
of orientation, to first order in the acceleration. We will now present this explicitly.


We begin by writing down the field of a charge undergoing hyperbolic motion with
constant proper acceleration $a_0$, as obsevered in the inertial (Minkowski) frame
in which the particle is momentarily at rest. If the particle is at $(X_0,y_0,z_0)$ then the
field at $(X,y,z)$ is given by  \cite{60Fulton,00EriksenI,12Steane}:
\be
\bar{E}_\rho &=& \frac{q}{r^3}\frac{(\rho-\rho_0) (1 + a_0 (X-X_0))}
{\left(1 + a_0(X-X_0) + a_0^2 r^2/4\right)^{3/2}}     \\
\bar{E}_x &=& \frac{q}{r^3} \frac{ (X\!-\!X_0) + \frac{a_0}{2}\left( (X\!-\!X_0)^2 - (\rho-\rho_0)^2\right)}
{ \left(1 + a_0(X-X_0) + a_0^2 r^2/4 \right)^{3/2}}  \nonumber
\ee
where $(\rho-\rho_0) = ((y-y_0)^2 + (z-z_0)^2)^{1/2}$ and $r = ( (X-X_0)^2 + (\rho-\rho_0)^2)^{1/2}$,
and we adopted Gaussian electromagnetic units.

In order to make the comparison with Pinto's calculation straightforward,
introduce a change of coordinates to $(t,x,y,z)$ where $t = \theta/g$, $x=h-1/g$
and $g$ is a constant. In these coordinates the metric is
\be
{\rm d}s^2 = - (1+g  x)^2 \dt^2 + {\rm d}x^2 + {\rm d}y^2 + {\rm d}z^2.   \label{xmetric}
\ee
In a general (i.e. not Minkowski) frame, there is more than one way to define what may be called
`electric field'. One possible definition is the spatial part of the local
4-force per unit charge on a charged particle that is not moving relative to the frame. This is given by
\be
{\cal E}^i =  F^i_{\; \lambda} u^\lambda      \label{Pad1}
\ee
(c.f. (\ref{dpEM})) where $u^\lambda$ is the 4-velocity of the local observer fixed in the frame. 
By the argument before eqn (\ref{sameforce}), we have
\be
{\cal E}^i = \bar{E}^i.
\ee
Also, since at $T=0$ we have $x = X-1/g$, it follows that $(X-X_0) = (x-x_0)$ , so the `electric field' 
in the Rindler frame, as given by (\ref{Pad1}) is,
\be
{\cal E}_\rho &=& \frac{q}{r^3}\frac{(\rho-\rho_0) (1 + a_0 (x-x_0))}{\left(1 + a_0(x-x_0) + a_0^2 r^2/4\right)^{3/2}}, \nonumber \\
{\cal E}_x &=& \frac{q}{r^3} \frac{ (x\!-\!x_0) + \frac{a_0}{2}\left( (x\!-\!x_0)^2 - (\rho-\rho_0)^2\right)}
{ \left(1 + a_0(x-x_0) + a_0^2 r^2/4 \right)^{3/2}} .  \label{field1}
\ee
In order that the charge at $x_0$ is fixed in the Rindler frame, its proper acceleration must match that of the local
observer in the frame, so
\be
a_0 = \frac{g}{1 + g x_0}.          \label{aproper}
\ee
The field given by (\ref{field1}) 
is the field described in \cite{62Bradbury,63Rohrlich,04Eriksen}. If we use it to calculate self-force,
the results will agree with those found in the Minkowski frame, as already noted.

In the accelerating
frame, the most natural way to form a sum of forces acting at different positions is the
one given by (\ref{newtot}). That is, one chooses for the hyperplane $\chi + {\rm d}\chi$ the next hyperplane
of simultaneity as defined by the acclerating frame. There remains a choice to be made about which worldline
is the reference worldline, whose proper time is $\tau_c$. Previously we suggested that one might use
the centroid of the accelerating composite object, but in order to study dynamics more generally, one
requires a reference worldline that is independent of the objects under consideration.
Therefore one picks the worldline
of a point fixed in the frame. The most natural such point is one at which $g_{00}$ has the value $-1$, since
then $\dtau_c = \dt$. For the metric (\ref{xmetric}) this is the case for $x_c = 0$, and we have
\be
\dby{\tau_x}{\tau_c} = \sqrt{-g_{00}} = 1 + gx .
\ee
We use this in (\ref{newtot}). Thus we find that the force per unit charge that must be summed in order
to calculate the self-force is given by
\be
E^i = \sqrt{-g_{00}}  F^i_{\; \lambda} u^\lambda = F^i_{\; 0}   \label{deffield}
\ee
which for our case is
\be
E^i = (1 + g x) {\cal E}^i.        \label{myfield}
\ee
For the case of a single point charge, to first order in $g$ this is
\be
E_{\rho} &\simeq& \frac{q \rho}{r^3} \left( 1 + \frac{g}{2} x) \right), \nonumber \\
E_x & \simeq & \frac{q}{r^3} 
\left( x + \frac{g}{2} \left[ 2x_0(x\! - \! x_0) - (\rho \! - \! \rho_0)^2 \right] \right).
\ee
These equations agree with equations (23)-(25) of Pinto \cite{06Pinto}. 

Let $\vE(\vvr_0, \vvr)$ be the field at $\vvr$ due to a point charge at $\vvr_0$, as given by
substituting (\ref{field1}) into (\ref{myfield}).
Then the electromagnetic self-force of a dipole formed by a pair of small charged spheres centred
at $(x_A,y_A,z_A)$ and $(x_B,y_B,z_B)$ is (ignoring the force of each sphere on itself)
\be
\vf_{\rm self} = -q^2 \left(\vE(\vvr_A, \vvr_B) + \vE(\vvr_B, \vvr_A)\right).
\ee
This force is in the $x$-direction. To 4th order in $g$ we thus find
\be
f_{\rm self} = \frac{q^2}{d} \left[g + \frac{3 d^2 - \Delta x^2}{8}\left(-g^3 + (x_A+x_B)g^4\right) \right]
\label{fselfRind}
\ee
where $d = |\vvr_B - \vvr_A|$ and $\Delta x = x_B-x_A$.
This agrees with eqn (26) of \cite{06Pinto}. Note that the force is independent of orientation of the dipole
at first order in $g$, and the lowest order orientation-dependent term is $O(g^3)$, in agreement
with (\ref{acubed}).

This completes the calculation of the electromagnetic self-force, but not the calculation of the
total self-force, which must also include the effects of internal stress. However, the pressure 
in the rod varies as $1/(1+gx)$ when calculated in the inertial instantaneous rest frame, and therefore
it is uniform when calculated in the accelerating frame, so in the latter frame it does not contribute a net force
when integrated over the whole surface of the rod.

In order to compare (\ref{fselfRind}) with eqn (\ref{fselftotal}), one should divide (\ref{fselfRind}) by
$\sqrt{-g_{00}(x_0)} = 1 + g x_0$, where $x_0 = (x_A+x_B)/2$, and use (\ref{aproper}).
Thus one finds the lowest order term agrees exactly with the result for an accelerating dipole
observed by an inertial observer. Therefore the system satisfies the Equivalence Principle.

\subsection{Defining the electric field}

In the above we discussed two definitions of what may be called `electric field'.
One natural definition is to take 
the spatial part of the four-force per unit charge:
\be
{\cal E}^i = u^\lambda F^i_{\; \lambda};   \;\;\;\; {\cal B}^i = \frac{1}{2} \epsilon^i_{\; \lambda\mu\nu} u^\lambda F^{\mu\nu}. \label{Pad}
\ee
This is recommended by Padmanabhan and Padmanabhan \cite{09Padmanabhan}, but, as we have discussed,
the notion of what is observed by observers fixed in the frame is better captured by including the
metric in the definition, so that one obtains (\ref{deffield}). This is the definition recommended by
Landau and Lifshitz \cite{71Landau}.
In the case of the Rindler frame (but not in general), the definition (\ref{Pad}) has
the following desirable feature: the field thus defined
in the Rindler frame is equal to that observed in the instantaneous inertial
rest frame of the local observer fixed in the Rindler frame. Since all observers fixed
anywhere in the Rindler frame share the same instantaneous inertial
rest frame, up to rotations 
(a special feature of certain frames, such as the Rindler frame), it follows that
the electric field $\cal E$ in the Rindler frame will be independent of a translation
of the coordinate system (a property not shared by $F^i_{\; 0}$). This was
noted in \cite{09Padmanabhan}; we have merely made an 
observation that allows it to be seen easily. Nevertheless, the field $F^i_{\; 0}$ is
the one best suited to examining what is observed by observers fixed in the frame,
especially when comparing or summing forces observed at different locations.

\section{Conclusion}  \label{s.conc}

To sum up, in this paper we have considered the electromagnetic self-force of
the electric dipole. 
It is a mistake to treat a dipole as a pair of point-like charged
particles of finite charge, because this amounts to assuming that the object under
discussion is equivalent to
an unphysical one, namely one with negative bare mass. Therefore one must consider
something more realistic. 
An object that is physically possible, and which approximates to an electric dipole, is
a pair of charged spherical shells of small radius $R$ whose centres are a small
distance $d$ apart, moving rigidly (i.e. with fixed proper size and shape).

We first examined the supposed self-accelerating dipole. We concluded that
the self-accelerating solution to the equation of motion is unphysical, because it
is based on the assumption that a physical object could have negative bare mass,
but that is not allowed in classical physics. 
In order to calculate this correctly, one must pay attention to
all the terms, including the inertial term in the self-force of the charged spheres.

The above conclusion was obtained analytically for the case $d \gg R$, and then extended to
all values of $d$ by performing a numerical integration. We find that
the total electromagnetic self-force is never along the direction of acceleration, 
for rigid hyperbolic motion of this system, and it vanishes
in the limit $d \rightarrow 0$ (for any fixed value of $R$). 

We also resolved a problem in relating the self-force to the expected inertia, when one compares
the cases of transverse and longitudinal acceleration. We argued that it turns on the inertia of 
pressure (or equivalently,
on the presence of hidden momentum), much like the famous `4/3 problem' for the charged sphere. 
In other words, one must include the effects of internal stresses in the
physical object under discussion. 
The new feature is that the charge distribution is not spherically symmetric so neither is the stress-energy
tensor. Hence the contribution of the internal stresses depends on the orientation of the system
relative to its acceleration. In general, the electromagnetic self-force of a physical object
{\em can} depend on the orientation of the object relative to its acceleration (and it does so depend for an
accelerating dipole), but the rate of change of `hidden momentum' in the object also has
such a dependence, with the net result that, for an isolated system in internal mechanical equilibrium,
the ratio of momentum to velocity is independent of orientation, to first order in
the acceleration, and is consistent with
the mass-energy equivalence, as required by Special Relativity.

We then noted that self-force is open to more  than one definition
(eqns (\ref{dptotA}) and (\ref{newtot})). This means that work based on the
second definition \cite{82Pearle,03Ori,04Ori} does not necessarily 
invalidate work based on the first, but in both cases one must pay attention to all
the relevant forces.

We next considered the effects of gravity.
We expect, of course, that the Principle of Equivalence will be upheld in any correct
General Relativistic treatment, but it is well known that that Principle needs careful handling
where self-force is concerned. It is useful to get precise algebraic statements
of what can and cannot be said about self-force in a set of
scenarios. We showed that an observer at rest in a gravitational
field described everywhere by the Rindler metric will find any charged object supported in the field
to possess an electromagnetic self-force equal to that observed in an inertial frame
when the same object moves with constant acceleration and fixed proper size
and shape. This is an exact statement about any charge distribution (not just a dipole or a sphere).

We showed how a recent calculation \cite{06Pinto} of 
the fields of a point charge in the Rindler metric may be reconciled with several earlier calculations \cite{62Bradbury,63Rohrlich,04Eriksen}, and \cite{09Padmanabhan}. We then used this
to obtain the self-force of a constantly accelerating dipole, as observed in the constantly
accelerating reference frame in which the dipole is at rest.

All the effects described in this paper can be explored, in principle, through sensitive mass
measurements of polar molecules.

I thank an anonymous referee who drew my attention to references \cite{03Ori,04Ori};
this facilitated an improved discussion of the gravitational effects.

\appendix
\section{Poincar\'e stresses}  \label{append1}
The term given by Eq. (\ref{poincare}) has been considered by many authors, starting with Henri Poincar\'e
in 1906 \cite{06Poincare,83Schwinger,92Yaghjian,06Medina}; for a brief
revue see Rohrlich \cite{97Rohrlich}. A simple way to calculate it
is as follows. First consider a charged sphere in inertial motion. 
We assume the charge is all situated in a thin shell on the surface of the sphere. In
the rest frame, the tension in the field at any point on the outer surface of this shell is 
\[
t = \frac{\epsilon_0}{2} \left(\frac{q}{4\pi \epsilon_0 R^2} \right)^2 = \frac{m_{\rm es}c^2}{4\pi R^3}. 
\]
Since the field inside the shell is zero, this is also the 
electromagnetic force per unit area on the shell of charge, in a radially outward direction
(one can also obtain it by arguing that each element of charge experiences an average field
equal to half that just outside the shell). For mechanical equilibrium, the material
of the sphere must provide a compensating inward force. We can most simply model this by
treating the sphere as an `ideal fluid', that is,
a continuous system which has a rest frame in which there is no sheer-stress, only pressure 
(or tension which is negative pressure). Then, for mechanical equilibrium, the pressure inside the sphere
must be equal to $-t$. In the relativistic equations of motion for an ideal fluid, the
energy density $\rho_0 c^2$ always enters in company with the pressure $p$, forming the combination
$(\rho_0 c^2 + p)$ \footnote{Eq. (16.24) of \cite{12Steane}.}. Consequently the
inertia of a fluid is modified by its pressure. For inertial motion, mechanical equilibrium is
attained if the pressure
is uniform throughout the volume of the sphere. By 
integrating $p/c^2$ over this volume one finds that 
the inertial mass of the sphere is modified by 
\[
\frac{4}{3}\pi R^3 p/c^2 = -\frac{4}{3} \pi R^2 t/c^2 = -\frac{1}{3}m_{\rm es}.
\]
When the sphere accelerates, the tension in the field changes somewhat, and the tension
in the sphere is no longer uniform. However, such modifications are of higher
than zeroth order in $R$. Therefore the above mass-modification, multiplied by the acceleration,
gives the leading order
contribution to the self-force owing to Poincar\'e stresses, as given in Eq. (\ref{poincare}).

\section{Numerical calculations}  \label{append2}

We wish to calculate the self-force for a `dipole' consisting of two rigid spherical charged shells of radius $R$ with
centres separated by $d$ and moving with constant proper acceleration in the transverse direction.
Such an entity has an instantaneous rest frame. In this frame, let $\vf^{(i,j)}$ be the net force on sphere $i$
owing to the electric field sourced by sphere $j$. Then, owing to the linearity of electromagnetism, the total self-force is
\[
\vf^{(1,1)} + \vf^{(1,2)} + \vf^{(2,1)} + \vf^{(2,2)} = 2 (f_{\rm shell} + f_{\rm dip}) \hat{\vx}
\]
where $f_{\rm shell}$ is given by Eq. (\ref{fselfexact}) and $f_{\rm dip}$ is equal to the $x$-component of
$\vf^{(2,1)}$. 

Choose the origin of coordinates so that the first sphere is centred at $(x,y,z)=(L_0,0,0)$ and the second
at $(L_0, d, 0)$, in the instantaneous rest frame, and both are accelerating in the positive $x$-direction.
Then
\be
f_{\rm dip} = \int E_x^{(1)}(\vvr_2) \, {\rm d}q_2
\ee
where $E_x^{(1)}(\vvr_2)$ is the $x$-component of the
electric field due to the first sphere at the location $\vvr_2$ of a point
on the second sphere, and ${\rm d}q_2 = (-q/4\pi R){\rm d}y_2 {\rm d}\phi_2$ is
an element of charge on the second sphere, in which $\phi_2$ is an azimuthal angle
about an axis through the centres of the spheres. $\phi_2$ is
in the range $0$ to $2\pi$, and $y_2$ ranges from $d-R$ to $d+R$.
In the overall rectangular coordinate system, such an element is located at
$(x_2,y_2,z_2)$ given by
\begin{eqnarray*}
x_2 &=& L_0 + \sqrt{R^2 - (y_2-d)^2} \, \cos(\phi_2), \\
y_2 &=& y_2, \\
z_2 &=& \sqrt{R^2 - (y_2-d)^2} \, \sin(\phi_2),
\end{eqnarray*}
where the positive square root should be taken.
As explained in \cite{13Steane}, to treat motion where the charge distribution undergoes
acceleration at fixed proper dimensions, the electric field in the integrand is given by
\be
E_x^{(1)}(\vvr_2) &=& \frac{q}{(4 \pi)^2 \epsilon_0 R} \int_{L_0-R}^{L_0+R} {\rm d}x_1
\int_0^{2\pi} {\rm d}\phi_1  \{ \nonumber \\
\lefteqn{  \tilde{E_x}\left(x_1;\; x_2, \;y_2-y_1,\; z_2-z_1 \right)   \}    }
\ee
where
\be
\tilde{E}_x(L;x,y,z) &\equiv& \frac{-4 L^2(L^2+y^2+z^2-x^2)}
{ \left( (L^2+x^2+y^2+z^2)^2 - 4 L^2 x^2 \right)^{3/2} },  \nonumber \\
y_1 &=& \sqrt{R^2 - (x_1-L_0)^2} \, \cos(\phi_1),  \nonumber  \\
z_1 &=& \sqrt{R^2 - (x_1-L_0)^2} \, \sin(\phi_1).
\ee
In order to handle the discontinuity in $E_x$, we used the `trick' described in section
\ref{s.numerical}. That is, before carrying out the integral to obtain $f_{\rm dip}$
we subtracted from $E_x$ a field with the same discontinuity and
whose contribution to the integral was zero.


\bibliography{selfforcerefs}

\begin{thebibliography}{34}
\expandafter\ifx\csname natexlab\endcsname\relax\def\natexlab#1{#1}\fi
\expandafter\ifx\csname bibnamefont\endcsname\relax
  \def\bibnamefont#1{#1}\fi
\expandafter\ifx\csname bibfnamefont\endcsname\relax
  \def\bibfnamefont#1{#1}\fi
\expandafter\ifx\csname citenamefont\endcsname\relax
  \def\citenamefont#1{#1}\fi
\expandafter\ifx\csname url\endcsname\relax
  \def\url#1{\texttt{#1}}\fi
\expandafter\ifx\csname urlprefix\endcsname\relax\def\urlprefix{URL }\fi
\providecommand{\bibinfo}[2]{#2}
\providecommand{\eprint}[2][]{\url{#2}}

\bibitem[{\citenamefont{Cornish}(1986)}]{86Cornish}
\bibinfo{author}{\bibfnamefont{F.~H.~J.} \bibnamefont{Cornish}},
  \bibinfo{journal}{Am. J. Phys.} \textbf{\bibinfo{volume}{54}},
  \bibinfo{pages}{166} (\bibinfo{year}{1986}).

\bibitem[{\citenamefont{Griffiths}(1986)}]{86Griffiths}
\bibinfo{author}{\bibfnamefont{D.~J.} \bibnamefont{Griffiths}},
  \bibinfo{journal}{Am. J. Phys.} \textbf{\bibinfo{volume}{54}},
  \bibinfo{pages}{744} (\bibinfo{year}{1986}).

\bibitem[{\citenamefont{Petkov}(1999)}]{99Petkov}
\bibinfo{author}{\bibfnamefont{V.}~\bibnamefont{Petkov}}
  (\bibinfo{year}{1999}), \bibinfo{note}{arXiv:physics/9906059}.

\bibitem[{\citenamefont{Rohrlich}(1990)}]{90Rohrlich}
\bibinfo{author}{\bibfnamefont{F.}~\bibnamefont{Rohrlich}},
  \emph{\bibinfo{title}{Classical Charged Particles: 3rd ed.}}
  (\bibinfo{publisher}{Addison-Wesley}, \bibinfo{address}{Reading,
  Massachusetts}, \bibinfo{year}{1990}).

\bibitem[{\citenamefont{Jackson}(1998)}]{98Jackson}
\bibinfo{author}{\bibfnamefont{J.~D.} \bibnamefont{Jackson}},
  \emph{\bibinfo{title}{Classical Electrodynamics}} (\bibinfo{publisher}{John
  Wiley}, \bibinfo{year}{1998}), \bibinfo{note}{3rd edition}.

\bibitem[{\citenamefont{Landau and Lifshitz}(1971)}]{71Landau}
\bibinfo{author}{\bibfnamefont{L.~D.} \bibnamefont{Landau}} \bibnamefont{and}
  \bibinfo{author}{\bibfnamefont{E.~M.} \bibnamefont{Lifshitz}},
  \emph{\bibinfo{title}{The Classical Theory of Fields}}
  (\bibinfo{publisher}{Pergamon}, \bibinfo{address}{Oxford},
  \bibinfo{year}{1971}), \bibinfo{note}{(1st edition (Russian) 1941)}.

\bibitem[{\citenamefont{Rohrlich}(1997)}]{97Rohrlich}
\bibinfo{author}{\bibfnamefont{F.}~\bibnamefont{Rohrlich}},
  \bibinfo{journal}{Am. J. Phys.} \textbf{\bibinfo{volume}{65}},
  \bibinfo{pages}{1051} (\bibinfo{year}{1997}).

\bibitem[{\citenamefont{Spohn}(2004)}]{04Spohn}
\bibinfo{author}{\bibfnamefont{H.}~\bibnamefont{Spohn}},
  \emph{\bibinfo{title}{Dynamics of charged particles and their radiation
  field}} (\bibinfo{publisher}{Cambridge U.P.}, \bibinfo{address}{Cambridge},
  \bibinfo{year}{2004}), \bibinfo{note}{arXiv:math-ph/9908024}.

\bibitem[{\citenamefont{Schwinger}(1983)}]{83Schwinger}
\bibinfo{author}{\bibfnamefont{J.}~\bibnamefont{Schwinger}},
  \bibinfo{journal}{Found. Phys.} \textbf{\bibinfo{volume}{13}},
  \bibinfo{pages}{373} (\bibinfo{year}{1983}).

\bibitem[{\citenamefont{Medina}(2006)}]{06Medina}
\bibinfo{author}{\bibfnamefont{R.}~\bibnamefont{Medina}},
  \bibinfo{journal}{J.Phys. A} \textbf{\bibinfo{volume}{39}},
  \bibinfo{pages}{3801} (\bibinfo{year}{2006}),
  \bibinfo{note}{arXiv:physics/0508031}.

\bibitem[{\citenamefont{Griffiths et~al.}(2010)\citenamefont{Griffiths,
  Proctor, and Schroeter}}]{10Griffiths}
\bibinfo{author}{\bibfnamefont{D.~J.} \bibnamefont{Griffiths}},
  \bibinfo{author}{\bibfnamefont{T.~C.} \bibnamefont{Proctor}},
  \bibnamefont{and} \bibinfo{author}{\bibfnamefont{D.~F.}
  \bibnamefont{Schroeter}}, \bibinfo{journal}{American Journal of Physics}
  \textbf{\bibinfo{volume}{78}}, \bibinfo{pages}{391} (\bibinfo{year}{2010}),
  \urlprefix\url{http://link.aip.org/link/?AJP/78/391/1}.

\bibitem[{\citenamefont{Griffiths and Owen}(1983)}]{83Griffiths}
\bibinfo{author}{\bibfnamefont{D.~J.} \bibnamefont{Griffiths}}
  \bibnamefont{and} \bibinfo{author}{\bibfnamefont{R.~E.} \bibnamefont{Owen}},
  \bibinfo{journal}{Am. J. Phys.} \textbf{\bibinfo{volume}{51}},
  \bibinfo{pages}{1120} (\bibinfo{year}{1983}).

\bibitem[{\citenamefont{Pinto}(2006)}]{06Pinto}
\bibinfo{author}{\bibfnamefont{F.}~\bibnamefont{Pinto}},
  \bibinfo{journal}{Phys. Rev. D} \textbf{\bibinfo{volume}{73}},
  \bibinfo{pages}{104020} (\bibinfo{year}{2006}),
  \urlprefix\url{http://link.aps.org/doi/10.1103/PhysRevD.73.104020}.

\bibitem[{\citenamefont{Ori and Rosenthal}(2003)}]{03Ori}
\bibinfo{author}{\bibfnamefont{A.}~\bibnamefont{Ori}} \bibnamefont{and}
  \bibinfo{author}{\bibfnamefont{E.}~\bibnamefont{Rosenthal}},
  \bibinfo{journal}{Phys. Rev. D} \textbf{\bibinfo{volume}{68}},
  \bibinfo{pages}{041701} (\bibinfo{year}{2003}),
  \bibinfo{note}{arXiv:gr-qc/020500},
  \urlprefix\url{http://link.aps.org/doi/10.1103/PhysRevD.68.041701}.

\bibitem[{\citenamefont{Ori and Rosenthal}(2004)}]{04Ori}
\bibinfo{author}{\bibfnamefont{A.}~\bibnamefont{Ori}} \bibnamefont{and}
  \bibinfo{author}{\bibfnamefont{E.}~\bibnamefont{Rosenthal}},
  \bibinfo{journal}{J. Math. Phys.} \textbf{\bibinfo{volume}{45}},
  \bibinfo{pages}{2347} (\bibinfo{year}{2004}),
  \bibinfo{note}{arXiv:gr-qc/0309102}.

\bibitem[{\citenamefont{Pearle}(1982)}]{82Pearle}
\bibinfo{author}{\bibfnamefont{P.}~\bibnamefont{Pearle}}, in
  \emph{\bibinfo{booktitle}{Electromagnetism: Paths to Research}}, edited by
  \bibinfo{editor}{\bibfnamefont{D.}~\bibnamefont{Teplitz}}
  (\bibinfo{address}{New York}, \bibinfo{year}{1982}), p. \bibinfo{pages}{211}.

\bibitem[{\citenamefont{Bradbury}(1962)}]{62Bradbury}
\bibinfo{author}{\bibfnamefont{T.~C.} \bibnamefont{Bradbury}},
  \bibinfo{journal}{Annals of Physics} \textbf{\bibinfo{volume}{19}},
  \bibinfo{pages}{323} (\bibinfo{year}{1962}).

\bibitem[{\citenamefont{Rohrlich}(1963)}]{63Rohrlich}
\bibinfo{author}{\bibfnamefont{F.}~\bibnamefont{Rohrlich}},
  \bibinfo{journal}{Ann. Phys} \textbf{\bibinfo{volume}{22}},
  \bibinfo{pages}{169} (\bibinfo{year}{1963}).

\bibitem[{\citenamefont{Eriksen and Gr\/on}(2004)}]{04Eriksen}
\bibinfo{author}{\bibfnamefont{E.}~\bibnamefont{Eriksen}} \bibnamefont{and}
  \bibinfo{author}{\bibfnamefont{O.}~\bibnamefont{Gr\/on}},
  \bibinfo{journal}{Annals of Physics} \textbf{\bibinfo{volume}{313}},
  \bibinfo{pages}{147} (\bibinfo{year}{2004}).

\bibitem[{\citenamefont{Fulton and Rohrlich}(1960)}]{60Fulton}
\bibinfo{author}{\bibfnamefont{T.}~\bibnamefont{Fulton}} \bibnamefont{and}
  \bibinfo{author}{\bibfnamefont{F.}~\bibnamefont{Rohrlich}},
  \bibinfo{journal}{Annals of Physics} \textbf{\bibinfo{volume}{9}},
  \bibinfo{pages}{499} (\bibinfo{year}{1960}).

\bibitem[{\citenamefont{Steane}(2012)}]{12Steane}
\bibinfo{author}{\bibfnamefont{A.~M.} \bibnamefont{Steane}},
  \emph{\bibinfo{title}{Relativity made relatively easy}}
  (\bibinfo{publisher}{Oxford U.P.}, \bibinfo{address}{Oxford},
  \bibinfo{year}{2012}).

\bibitem[{\citenamefont{Steane}(2013)}]{13Steane}
\bibinfo{author}{\bibfnamefont{A.~M.} \bibnamefont{Steane}},
  \bibinfo{journal}{Proc. R. Soc. A} \textbf{\bibinfo{volume}{470}},
  \bibinfo{pages}{20130480} (\bibinfo{year}{2013}),
  \bibinfo{note}{arXiv:physics:1307.5011}.

\bibitem[{\citenamefont{Roa-Neri and Jimenez}(1994)}]{94RoaNeri}
\bibinfo{author}{\bibfnamefont{J.~A.~E.} \bibnamefont{Roa-Neri}}
  \bibnamefont{and} \bibinfo{author}{\bibfnamefont{J.~L.}
  \bibnamefont{Jimenez}}, \bibinfo{journal}{Foundations of Physics Letters}
  \textbf{\bibinfo{volume}{7}}, \bibinfo{pages}{403} (\bibinfo{year}{1994}).

\bibitem[{\citenamefont{Page and Adams}(1945)}]{45Page}
\bibinfo{author}{\bibfnamefont{L.}~\bibnamefont{Page}} \bibnamefont{and}
  \bibinfo{author}{\bibfnamefont{N.~I.} \bibnamefont{Adams}},
  \bibinfo{journal}{Am. J. Phys.} \textbf{\bibinfo{volume}{13}},
  \bibinfo{pages}{141} (\bibinfo{year}{1945}).

\bibitem[{\citenamefont{Rainville et~al.}(2004)\citenamefont{Rainville,
  Thompson, and Pritchard}}]{04Rainville}
\bibinfo{author}{\bibfnamefont{S.}~\bibnamefont{Rainville}},
  \bibinfo{author}{\bibfnamefont{J.~K.} \bibnamefont{Thompson}},
  \bibnamefont{and} \bibinfo{author}{\bibfnamefont{D.~E.}
  \bibnamefont{Pritchard}}, \bibinfo{journal}{Science}
  \textbf{\bibinfo{volume}{303}}, \bibinfo{pages}{334} (\bibinfo{year}{2004}).

\bibitem[{\citenamefont{Poisson et~al.}(2011)\citenamefont{Poisson, Pound, and
  Vega}}]{11Poisson}
\bibinfo{author}{\bibfnamefont{E.}~\bibnamefont{Poisson}},
  \bibinfo{author}{\bibfnamefont{A.}~\bibnamefont{Pound}}, \bibnamefont{and}
  \bibinfo{author}{\bibfnamefont{I.}~\bibnamefont{Vega}},
  \bibinfo{journal}{Living Rev. Rel.} \textbf{\bibinfo{volume}{14}},
  \bibinfo{pages}{7} (\bibinfo{year}{2011}), \bibinfo{note}{arXiv:1102.0529}.

\bibitem[{\citenamefont{Soichiro~Isoyama}(2012)}]{12Isoyama}
\bibinfo{author}{\bibfnamefont{E.~P.} \bibnamefont{Soichiro~Isoyama}}
  (\bibinfo{year}{2012}), \bibinfo{note}{arXiv:1205.1236 [gr-qc]}.

\bibitem[{\citenamefont{Vilenkin}(1979)}]{79Vilenkin}
\bibinfo{author}{\bibfnamefont{A.}~\bibnamefont{Vilenkin}},
  \bibinfo{journal}{Phys. Rev. D} \textbf{\bibinfo{volume}{20}},
  \bibinfo{pages}{373} (\bibinfo{year}{1979}).

\bibitem[{\citenamefont{Padmanabhan and Padmanabhan}(2009)}]{09Padmanabhan}
\bibinfo{author}{\bibfnamefont{H.}~\bibnamefont{Padmanabhan}} \bibnamefont{and}
  \bibinfo{author}{\bibfnamefont{T.}~\bibnamefont{Padmanabhan}}
  (\bibinfo{year}{2009}), \bibinfo{note}{arXiv:0910.0926 [gr-qc]}.

\bibitem[{\citenamefont{Rindler}(2006)}]{06Rindler}
\bibinfo{author}{\bibfnamefont{W.}~\bibnamefont{Rindler}},
  \emph{\bibinfo{title}{Relativity: Special, General, and Cosmological: 2nd
  ed.}} (\bibinfo{publisher}{Oxford U.P.}, \bibinfo{address}{Oxford},
  \bibinfo{year}{2006}).

\bibitem[{\citenamefont{Eriksen and Gr\/on}(2000)}]{00EriksenI}
\bibinfo{author}{\bibfnamefont{E.}~\bibnamefont{Eriksen}} \bibnamefont{and}
  \bibinfo{author}{\bibfnamefont{O.}~\bibnamefont{Gr\/on}},
  \bibinfo{journal}{Annals of Physics} \textbf{\bibinfo{volume}{286}},
  \bibinfo{pages}{320} (\bibinfo{year}{2000}).

\bibitem[{\citenamefont{Poincar\'e}(1906)}]{06Poincare}
\bibinfo{author}{\bibfnamefont{H.}~\bibnamefont{Poincar\'e}},
  \bibinfo{journal}{Rendiconti del Circolo Matematico di Palermo}
  \textbf{\bibinfo{volume}{21}}, \bibinfo{pages}{129} (\bibinfo{year}{1906}).

\bibitem[{\citenamefont{Yaghjian}(1992, 2006)}]{92Yaghjian}
\bibinfo{author}{\bibfnamefont{A.~D.} \bibnamefont{Yaghjian}},
  \emph{\bibinfo{title}{Relativistic Dynamics of a Charged Sphere: 2nd ed.}}
  (\bibinfo{publisher}{Springer-Verlag}, \bibinfo{address}{Berlin},
  \bibinfo{year}{1992, 2006}).

\bibitem[{\citenamefont{Griffiths and Szeto}(1978)}]{78Griffiths}
\bibinfo{author}{\bibfnamefont{D.~J.} \bibnamefont{Griffiths}}
  \bibnamefont{and} \bibinfo{author}{\bibfnamefont{E.~W.} \bibnamefont{Szeto}},
  \bibinfo{journal}{Am. J. Phys.} \textbf{\bibinfo{volume}{46}},
  \bibinfo{pages}{244} (\bibinfo{year}{1978}).

\end{thebibliography}

\end{document}